\documentclass[aps,prl,twocolumn,superscriptaddress]{revtex4}

\usepackage{graphicx}
\usepackage{dcolumn}
\usepackage{bm}
\begin{document}

\preprint{APS/123-QED}

\title{Coherent quasi-particles-to-incoherent hole-carriers crossover in underdoped cuprates}

\author{M. Hashimoto}
\affiliation{Department of Physics, University of Tokyo, Tokyo 113-8656, Japan}
\author{T. Yoshida}
\affiliation{Department of Physics, University of Tokyo, Tokyo 113-8656, Japan}
\author{K. Tanaka}
\affiliation{Department of Physics, University of Tokyo, Tokyo 113-8656, Japan}
\author{A. Fujimori}
\affiliation{Department of Physics, University of Tokyo, Tokyo 113-8656, Japan}
\author{M. Okusawa}
\affiliation{Department of Physics, Faculty of Education, Gunma University, Maebashi, Gunma 371-8510, Japan}
\author{S. Wakimoto}
\affiliation{QuBS, Japan Atomic Energy Agency, Tokai, Ibaraki, 319-1195, Japan}
\author{K. Yamada}
\affiliation{Institute of Materials Research, Tohoku University, Sendai 980-8577, Japan}
\author{T. Kakeshita}
\affiliation{SRL- ISTEC, Tokyo, 135-0062, Japan}
\author{H. Eisaki}
\affiliation{AIST, 1-1-1 Central 2, Umezono, Tsukuba, Ibaraki, 305-8568, Japan}
\author{S. Uchida}
\affiliation{Department of Physics, University of Tokyo, Tokyo 113-8656, Japan}

\date{\today}

\begin{abstract}
In underdoped cuprates, only a portion of the Fermi surface survives as Fermi arcs due to pseudogap opening. 
In hole-doped La$_{2}$CuO$_4$, we have deduced the ``coherence temperature'' $T_{coh}$ of quasi-particles on the Fermi arc above which the broadened leading edge position in angle-integrated photoemission spectra is shifted away from the Fermi level and the quasi-particle concept starts to lose its meaning.
$T_{coh}$ is found to rapidly increase with hole doping, an opposite behavior to  the pseudogap temperature $T^*$. 
The superconducting dome is thus located below both $T^*$ and $T_{coh}$, indicating that the superconductivity emerges out of the coherent Fermionic quasi-particles on the Fermi arc.
$T_{coh}$ remains small in the underdoped region, indicating that incoherent charge carriers originating from the Fermi arc are responsible for the apparently metallic transport at high temperatures. 
\end{abstract}


\maketitle
In order to understand the variety of interesting phenomena of doped Mott insulators \cite{Pollini01,Fujimori1995}, it is necessary to reveal how the electrons behave as a function of carrier concentration \cite{Mizutani2001}.
High-$T_c$ superconductivity in the cuprates is one of the most spectacular examples of doped Mott insulators in which superconductivity \cite{Tinkham1996} emerges from an unconventional normal state. 
Photoemission studies have revealed a wealth of its electronic structure \cite{DamascelliAngle-resolved03}.
Recent angle-resolved photoemission (ARPES) studies of underdoped cuprates have revealed that the Fermi arc in the nodal region, which remains without a pseudogap above $T_c$ and governs the normal-state transport, plays important roles on the superconductivity, too \cite{TanakaDistinct06, LeeAbrupt07,KondoEvidence07, YoshidaLow-energy07, KanigelEvolution06}.
Below $T_c$, a $d$-wave superconducting gap opens on the Fermi arc, while the pseudogap in the antinodal region, which opens below pseudogap temperature $T^*$ ($>T_c$), does not show a strong temperature dependence across $T_c$.
Thus, it has been suggested that the high-$T_c$ superconductivity is realized on the Fermi arc and that the pseudogap is not directly connected to the superconductivity \cite{TanakaDistinct06, LeeAbrupt07,KondoEvidence07, HashimotoDistinct07}.
It has been also reported that the Fermi arc shrinks with underdoping \cite{LeeAbrupt07, YoshidaLow-energy07, KanigelEvolution06} corresponding to the carrier number which decreases like $n\sim x$ and the arc length becomes longer with temperature \cite{KanigelEvolution06} until the Fermi surface recovers above $T^*$.

On the other hand, it has not been obvious how the quasi-particles on the Fermi arc change with temperature as well as doping.  
If one defines the Fermi energy $\epsilon _F\propto n/m^*$ of the doped holes, it should decrease with underdoping, since the carrier mobility  $\mu$ $\propto 1/m^*$ decreases only slowly \cite{AndoMobility01}. 
As the temperature increases from below $T_F\equiv \epsilon _F/k_B$ to above it, the doped holes would change their character from a degenerate Fermi liquid (on the Fermi arc) consisting of coherent quasi-particles obeying the Fermi statistics to a classical gas of (incoherent) holes obeying the Boltzmann statistics.
Therefore, if $T_F < T^*$, it is expected that charge carriers will lose its quasi-particle properties before the entire Fermi surface is recovered by the collapse of the pseudogap.
So far, there has not been a quantitative experimental estimate of $T_F$ in the under doped cuprates.
Therefore, in order to observe such a crossover, we have performed systematic temperature and doping dependent angle-integrated photoemission (AIPES) measurements of the single-layer cuprates La$_{2-x}$Sr$_x$CuO$_{4}$ (LSCO) and La$_{2}$CuO$_{4.10}$, and derived the crossover temperature or the ``coherence temperature'' $T_{coh}$, which should essentially follow $T_F$.

\begin{figure*}
\begin{center}
\includegraphics[width=\linewidth]{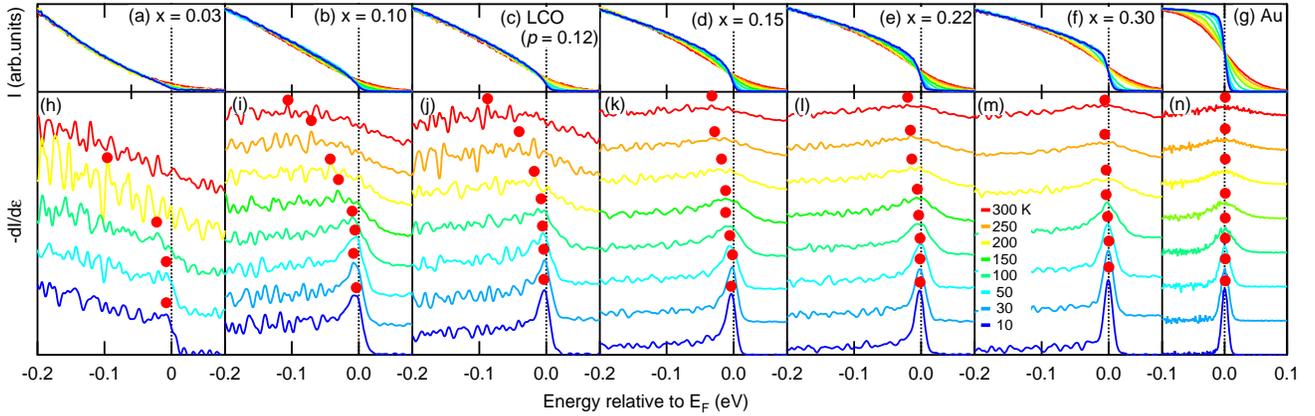}
\caption{(Color online) Doping and temperature dependences of the AIPES spectra near \textit{E$_F$} for La$_{2-x}$Sr$_x$CuO$_{4}$ (LSCO), La$_2$CuO$_{4.10}$ (LCO, $p$ $\sim$ 0.12) and gold reference.
(a)-(g) Raw spectra reproduced from \cite{HashimotoDistinct07}.
(h)-(n) First derivative curves of the AIPES spectra.
Each spectrum is shifted vertically so that one can see the peaks clearly.
Red symbols show the peak positions.
}

\label{IntD}
\end{center}
\end{figure*}

We measured LSCO samples with $x$ = 0.03, 0.10 (\textit{T$_c$} = 25 K), 0.15 (\textit{T$_c$} = 38 K), 0.22 (\textit{T$_c$} = 28 K), 0.30, and La$_2$CuO$_{4.10}$ (LCO) with hole concentration \textit{p} $\sim$ 0.12 (\textit{T$_c$} $\sim$ 35 K).
The sample temperature was varied between 10 K and 300 K.
The total energy resolution was set at $\sim$10 meV.
Because of the high stability of the power supply of the analyzer, the accuracy in determining $E_F$ was within 1 meV.
Experimental details were described before \cite{HashimotoDistinct07}. 

In Fig. \ref{IntD}(a)-(g), we have reproduced the temperature dependent photoemission spectra of LSCO, LCO and gold near $E_F$ from Ref. \cite{HashimotoDistinct07}. 
One can see that the Fermi edge and its temperature dependence are most clearly observed for gold as well as in the overdoped samples, but becomes blurred with underdoping at elevated temperatures.
In the underdoped region, the edge becomes less clear with temperature than that in the overdoped region, indicating that the Fermi-Dirac statistics lose its meaning with temperature in the underdoped region.
In order to evaluate the disappearance of the Fermi edge quantitatively, that is, crossover from coherent quasi-particles to incoherent hole carriers, we have analyzed the spectra using first derivative and scaling relationship, as we shall describe below.

Figure \ref{IntD}(h)-(n) shows the (smoothed) first derivative curves of the spectra.
The peak positions indicated by red symbols can be regarded as the leading edge mid-point of the raw spectra.
For the gold spectra [Fig. \ref{IntD}(n)], the peak of the first derivative curves appears exactly at $E_F$ at any temperature although the peak becomes broader with temperature, as expected from the Fermi-Dirac (FD) distribution function.
In the case of the heavily overdoped LSCO of $x$ = 0.30 too, one can observe a peak up to 300 K as in the case of gold.
However, there is a slight shift of the peak position toward below $E_F$ and the peak shows an asymmetric tail extending below $E_F$, reflecting the strong slope of the density of states (DOS).
With decreasing hole concentration, the shift of the peak starts at a lower temperature and becomes stronger.
The asymmetric broadening of the peak with increasing temperature also becomes more pronounced.
For the most underdoped $x$ = 0.03 sample, only at low temperatures, one can barely see a small peak near $E_F$ arising from the tiny Fermi cut-off.
With increasing temperature, the peak is rapidly shifted away from $E_F$, and becomes ambiguous and difficult to define, corresponding to the disappearance of the Fermi edge in the raw spectra [Fig. \ref{IntD}(a)].

\begin{figure}
\begin{center}
\includegraphics[width=0.85\linewidth]{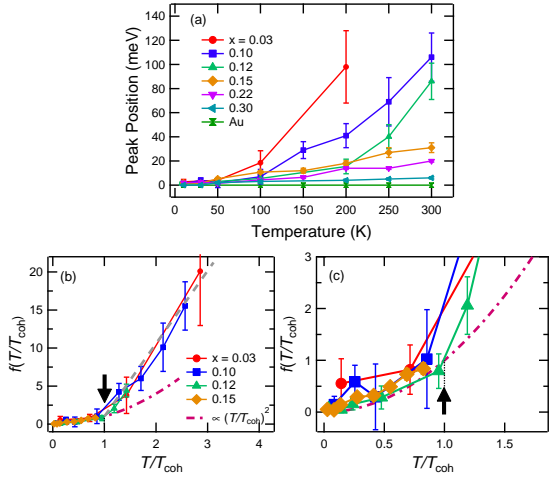}
\caption{(Color online) Temperature dependence of the peak position $E_{\mathrm{peak}}(x, T)$ in the first derivative of spectra for LSCO and LCO, together with gold reference.
(a) Raw peak position $E_{\mathrm{peak}}(x, T)$.
(b) Scaling plot of the peak position $E_{\mathrm{peak}}(x,T)/E_{coh}(x)=f(T/T_{coh}(x))$.
One can see that, for $T$ \textless $T_{coh}$, $f(T/T_{coh})\propto (T/T_{coh})^2$ and for $T$ \textgreater $T_{coh}$ is approximately linear.
Arrows indicate $T$ = $T_{coh}$ 
(c) Enlarged plot of (b)
}
\label{scaling}
\end{center}
\end{figure}

In Fig. \ref{scaling}(a), the peak position $E_{peak}(x,T)$ of the first derivative curves is plotted as a function of temperature for various hole concentrations.
With decreasing hole concentration, the deviation of the peak position from $E_F$ occurs at lower temperatures as mentioned above.
If one defines $T_{coh}$ by the temperature above which the deviation of the peak position from $E_F$ becomes significant, Fig. \ref{scaling}(a) indicates that $T_{coh}$ increases with decreasing hole concentration.
$T_{coh}$ may be regarded as the crossover temperature from the Fermi-liquid-like metallic state to a non-metallic (or incoherent-metallic) one which may be considered as a collection of hole carriers (polarons?) that show incoherent hopping. 
In order to deduce the $x$ dependence of the coherence temperature $T_{coh}(x)$ from the set of experimental data, we have performed a scaling analysis of the peak position $E_{\mathrm{peak}}(x, T)$.
By assuming that $E_{\mathrm{peak}}(x, T)$ obeys the scaling relation $E_{\mathrm{peak}}(x,T)/E_{coh}(x)=f(T/T_{coh}(x))$, where $E_{coh}(x)$ is the $x$-dependent ``coherence energy scale'' ($\sim$$\epsilon _F/k_B$ as we shall see below), all the data points fall onto a single curve as shown in Fig. \ref{scaling}(b) and (c) \cite{foot1}. 
In this plot, one can see a temperature-dependent crossover at $T/T_{coh}(x)$ $\sim$ 1, from the weakly temperature dependent $E_{coh}(x, T)$ to the strongly temperature dependent $E_{coh}(x, T)$.
In the case of the Fermi liquid, a simulation for a DOS with a finite slope multiplied by the FD function has shown that the leading edge show a small shift approximately proportional to $T^2$.
One can see that the small shifts for $T/T_{coh}$ \textless 1 are consistent with the $T^2$ behavior as shown in Fig. \ref{scaling}(c).
At higher temperatures $T/T_{coh}$ \textgreater 1, the leading edge becomes less well-defined and show more rapid shift than in the low-temperature region.
In fact, the peak shift is practically linear in $T$.
We note that a $T\ln T$ behavior is expected for the chemical potential of a classical hole gas obeying the Boltzmann statistics, which is almost linear in $T$.

\begin{figure}
\begin{center}
\includegraphics[width=0.8\linewidth]{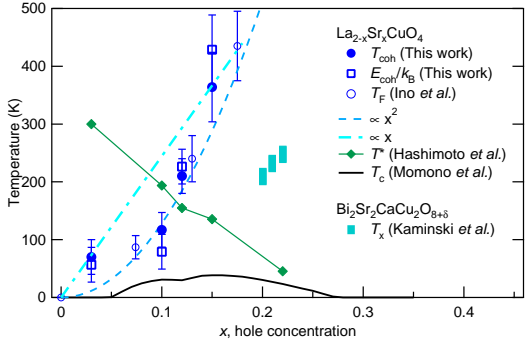}
\caption{(Color online) Characteristic temperatures for LSCO: $T_{coh}(x)$, $E_{coh}(x)/k_B$, $T_F$ (Ref. \cite{InoDoping98}), $T_c$ (Ref. \cite{MomonoCrossover99}) and $T^*$ (Ref. \cite{HashimotoDistinct07}).
Interlayer coherence temperature $T_x$ for Bi2212 (Ref. \cite{KaminskiCrossover03}) are plotted together.}
\label{PD}
\end{center}
\end{figure}

$T_{coh}(x)$ and $E_{coh}(x)$ thus deduced are plotted in Fig. \ref{PD}.
For $x$ = 0.03, $T_{coh}$ is lower than 100 K , but increases quickly with increasing hole concentration.
It exceeds 300 K for optimally doped $x$ = 0.15 and becomes even higher for $x$ = 0.22 and 0.30.
Note that $E_{coh}(x)$ and $T_{coh}(x)$ are mutually consistent as they satisfy $E_{coh}(x)$ $\sim$ $k_BT_{coh}(x)$.
The present results are in accordance with the previous work on LSCO where the hole Fermi energy $\epsilon_{F}$ is estimated at low temperatures assuming linearly decreasing model DOS reminiscent of a hole-doped semiconductor \cite{InoDoping98}.

Figure \ref{PD} also clearly shows that the doping dependence of $T_{coh}$ is opposite to that of the pseudogap temperature $T^*$.
One can also see that the superconductivity is realized below both $T_{coh}$ and $T^*$.
That is, only when the Fermi-Dirac statistics is valid and the quasi-particles become well defined ($T$ \textless $T_{coh}$) on the Fermi arc, the superconductivity appears.
Recently, there has been accumulating evidence that in the underdoped cuprates the superconductivity is realized mainly on the arc in the nodal region of the Fermi surface and that the $\sim$($\pi$, 0) region does not make significant contribution to the superconductivity \cite{TanakaDistinct06, LeeAbrupt07, TaconTwo06, HashimotoDistinct07,KondoEvidence07}.
In underdoped region $T_{coh}<T^*$, which means that with increasing temperature, the quasi-particles on the Fermi arc start to lose their coherence before the Fermi surface is recovered at $T^*$.
This means that, in the temperature range $T_{coh} < T < T^*$, the pseudogap opens on the ``incoherent'' portion of the Fermi surface.
The present finding of the pseudogap formed from incoherent carriers and the superconducting gap from coherent quasi-particles may be an important point to understand the different natures of the two gaps.
A temperature dependent ARPES study has reported that the loss of coherence in the antinodal region of the underdoped Bi2212 occurs around 150 - 200 K \cite{KordyukNonmonotonic09}, which is close to present estimation of $T_{coh}$, while another paper has indicated that the well defined quasiparticles in the nodal region survives at 130 K \cite{KordyukBare05}. The estimated $T_{coh}$ in the present paper may be understood as the temperature above which the incoherent carriers become dominant near $E_{F}$. The material dependence of $T_{coh}$ can be another possibility for those observations in Bi2212.

If we regard $k_BT_{coh}$ to be equal to the Fermi energy $\epsilon _F$ of the doped holes, it is natural that $T_{coh}$ increases with $x$. 
The small $\epsilon_F \sim k_BT_{coh}$ at low doping levels suggests that the top of the ``valence band'' into which holes are doped is very close to $E_F$.
This behavior is reminiscent of the small hole pocket, although a large Fermi surface truncated by the pseudogap into the Fermi arcs are seen by ARPES.
For the notion of the hole Fermi energy to be meaningful, the DOS above $E_F$ must decrease with energy as assumed in the linear model DOS \cite{InoDoping98}.
Such a decrease of the DOS above $E_F$ may be related to the well-known electron-hole asymmetry observed in STM data \cite{KohsakaIntrinsic07}.
The asymmetry reported by STM studies becomes strong with underdoping, consistent with the doping dependence of the slope of the DOS in the present AIPES spectra.

Here, it should be remarked that the interlayer coherence in the bilayer cuprate Bi2212 has been studied by ARPES in the overdoped region and the interlayer coherence temperature $T_x$ shows a qualitatively similar doping dependence to $T_{coh}$ \cite{KaminskiCrossover03} as plotted in Fig. \ref{PD}. [$T_x$ $\sim$ (1/3)$T_{coh}$]. 
While the intralayer coherence involves quasi-particles around the node, the interlayer coherence has been examined in the antinodal region where the bilayer splitting is maximum. 
$T_{coh} \gg T_x$ means that intralayer coherence is a necessary condition for the quasi-particles to make coherent hopping between the CuO$_2$ layers.  
We note that there is seemingly a contradicting report to $T_x$ that the disappearance of quasiparticles occurs at $T_c$ rather than at $T_x$ \cite{DingCoherent01}.

The doping dependence of $T_{coh}$ may need further consideration.
When holes are doped into a 2D antiferromagnetic insulator, the doping dependence of the Fermi energy $\epsilon _F$ ($\propto $$T_{coh}$) should be proportional to $x$.
According to Gutzwiller approximation \cite{Zhangrenormaised88}, the band width is predicted to be $2xt/(1+x)$, where $t$ is the nearest hopping parameter.
As shown in Fig. \ref{PD}, however, the present $T_{coh}(x)$ remains small (\textless100 K) below $x\sim0.1$ and quickly increases above it and appears to show a superlinear $x$ dependence rather than $x$-linear doping dependence.
In the case of the 2D doped Mott insulator, Imada \cite{ImadaMetal-Insulator95} has indicated that, near the metal-insulator transition (MIT), the doping dependence of $T_{coh}$ and chemical potential $\mu$ behave as $T_{coh}$ $\propto$ $\mu$ $\propto$ $x^{z/d}$ according to hyperscaling, where $z$ is the dynamical exponent of the MIT and $d$ is the spatial dimension ($d$ = 2).
Figure \ref{PD} shows that the doping dependence of $T_{coh}$ is more consistent with $\propto$ $x^2$ than $\propto$ $x$, implying that $z$ $\sim$ 4, which is different from that of a band insulator-to-metal transition, where $z$ = 2. 
Parcollet and Georges \cite{ParcolletNon-Fermi-liquid99} have also indicated the $x^2$ doping dependence of $T_{coh}$ in the low-doping regime of the $t-J$ model.
The $x^2$ doping dependence of $T_{coh}$ is also consistent with the suppression of the chemical potential shift in the underdoped region \cite{InoChemical97}, which also indicates an anomalously large exponent $z$ \cite{FurukawaTwo-Dimensional92}.

Finally, we compare the present result with the mobility $\mu$ of doped carriers reported by Ando $et$ $al$ \cite{AndoMobility01}.
While $\mu$ at a fixed temperature depends on doping, we find that $\mu$ at $T_{coh}$ is almost doping independent: 
For $x$ = 0.03, $\mu$ at $T_{coh}$ $\sim$ 75 K is $\sim$10 cm$^2$/Vs and for $x$ = 0.12, $\mu$ at $T_{coh}$ $\sim$ 210 K is $\sim$11.1 cm$^2$/Vs.
Here, $\mu\equiv \sigma/n$ has been estimated under the assumption that $n$ = $x$ \cite{AndoMobility01}.
This ``critical'' $\mu$ corresponds to the mean-free path of $\sim$ 30 $\mathrm{\AA}$, as large as several lattice constants.
This means that the incoherent carrier is not localized in a unit cell but is extended over several lattice constants. 
The coherence of charge carriers can also be monitored by the spectral weight around $\omega = 0$ or Drude weight in optical conductivity \cite{DummAnisotropic03, TakenakaCoherent-to-incoherent02, TakenakaIncoherent03, OrtolaniFrequency-Dependent05}.
Depression of the Drude weight with temperature has been observed as expected theoretically \cite{JakliifmmodeCharge95}. 
At a fixed low temperature, Drude weight \cite{DummAnisotropic03, TakenakaCoherent-to-incoherent02, TakenakaIncoherent03, OrtolaniFrequency-Dependent05}, the mobility of carriers \cite{AndoMobility01} and the nodal spectral weight at $E_F$ from ARPES measurements \cite{YoshidaMetallic03} increase with doping.
These doping dependences are consistent with the doping dependence of $T_{coh}$, although it is difficult to estimate the value of $T_{coh}$ from these measurements.

In conclusion, in the underdoped region, the quasi-particles on the Fermi arc start to lose its coherence quickly above $T_{coh}$, which exhibits opposite doping dependence to $T^*$.
The result that $T_c$ \textless $T_{coh}$ indicates that Cooper pairs are formed from quasi-particles, consistent with the picture that the superconductivity in the underdoped region is realized on the Fermi arc in the nodal region.
Whether there is correlation between $T_{coh}$ and $T_{c}$ would give insight into the mechanism of superconductivity.
It is also important that the apparently metallic transport at higher temperatures ($>T_{coh}$) in the underdoped region is governed by incoherent hole carriers on the Fermi arc/surface.
$T_{coh}$ increases more rapidly with doping than $\propto x$, consistent with the suppressed chemical potential shift near the filling-controlled Mott transition in the hyperscaling framework and reflects the peculiarity of the Mott insulator-to-metal transition in 2D.

Informative discussion with N. Bontemps and A.-F. Santander-Syro is gratefully acknowledged. 
This work was supported by a Grant-in-Aid for Scientific Research in Priority Area ``Invention of Anomalous Quantum Materials'' from MEXT, Japan.


\begin{thebibliography}{10}

\bibitem{Pollini01}
I. Pollini, A. Mosser, and J.~C. Parlebas, Physics Reports {\bf 355},  1
  (2001).

\bibitem{Fujimori1995}
A. Fujimori,  in {\em Strong Correlation and Superconductivity}, {\em Springer
  Series in Solid State Science119}, edited by A. Fujimori and Y. Tokura
  (Spring -- Verlag, Berlin, 1995).

\bibitem{Mizutani2001}
U. Mizutani, {\em Electron theory of metals} (Cambridge University Press,
  Cambridge, 2001).

\bibitem{Tinkham1996}
M. Tinkham, {\em Introduction to Superconductivity} (McGraw Hill Inc., New
  York, 1996).

\bibitem{DamascelliAngle-resolved03}
A. Damascelli, Z. Hussain, and Z.-X. Shen, Rev. Mod. Phys. {\bf 75},  473
  (2003).

\bibitem{TanakaDistinct06}
K. Tanaka {\it et~al.}, Science {\bf 314},  1910  (2006).

\bibitem{LeeAbrupt07}
W.~S. Lee {\it et~al.}, Nature {\bf 450},  81  (2007).

\bibitem{KondoEvidence07}
T. Kondo, T. Takeuchi, A. Kaminski, S. Tsuda and S. Shin, Phys. Rev. Lett. {\bf 98},  267004  (2007).

\bibitem{YoshidaLow-energy07}
T. Yoshida {\it et~al.}, J. Phys. Condens. Matter {\bf 19},  125209  (2007).

\bibitem{KanigelEvolution06}
A. Kanigel {\it et~al.}, Nature Phys. {\bf 2},  447  (2006).

\bibitem{HashimotoDistinct07}
M. Hashimoto {\it et~al.}, Phys. Rev. B {\bf 75},  140503(R)  (2007).

\bibitem{AndoMobility01}
Y. Ando, A.~N. Lavrov, S. Komiya, K. Segawa and X.~F. Sun, Phys. Rev. Lett. {\bf 87},  017001  (2001).

\bibitem{foot1}
T. effect of the finite resolution in the peak shift can be eliminated based on
  a simulation assuming a~linear DOS.

\bibitem{InoDoping98}
A. Ino {\it et~al.}, Phys. Rev. Lett. {\bf 81},  2124  (1998).

\bibitem{MomonoCrossover99}
N. Momono {\it et~al.}, Physica C {\bf 317},  603  (1999).

\bibitem{KaminskiCrossover03}
A. Kaminski {\it et~al.}, Phys. Rev. Lett. {\bf 90},  207003  (2003).

\bibitem{TaconTwo06}
M.~L. Tacon {\it et~al.}, Nature phys. {\bf 2},  573  (2006).

\bibitem{KordyukNonmonotonic09}
A.~A. Kordyuk {\it et~al.}, Phys. Rev. B {\bf 79},  020504(R)  (2009).

\bibitem{KordyukBare05}
A.~A. Kordyuk {\it et~al.}, Phys. Rev. B {\bf 71},  214513  (2005).

\bibitem{KohsakaIntrinsic07}
Y. Kohsaka {\it et~al.}, Science {\bf 315},  1380  (2007).

\bibitem{DingCoherent01}
H. Ding {\it et~al.}, Phys. Rev. Lett. {\bf 87},  227001  (2001).

\bibitem{Zhangrenormaised88}
F. Zhang, C. Gros, T. Rice, and H. Shiba, Supercond. Sci. Technol. {\bf 1},  36
   (1988).

\bibitem{ImadaMetal-Insulator95}
M. Imada, J. Phys. Soc. Jpn. {\bf 64},  2954  (1995).

\bibitem{ParcolletNon-Fermi-liquid99}
O. Parcollet and A. Georges, Phys. Rev. B {\bf 59},  5341  (1999).

\bibitem{InoChemical97}
A. Ino {\it et~al.}, Phys. Rev. Lett. {\bf 79},  2101  (1997).

\bibitem{FurukawaTwo-Dimensional92}
N. Furukawa and M. Imada, J. Phys. Soc. Jpn. {\bf 61},  3331  (1992).

\bibitem{DummAnisotropic03}
M. Dumm, S. Komiya, Y. Ando, and D.~N. Basov, Phys. Rev. Lett. {\bf 91},
  077004  (2003).

\bibitem{TakenakaCoherent-to-incoherent02}
K. Takenaka {\it et~al.}, Phys. Rev. B {\bf 65},  092405  (2002).

\bibitem{TakenakaIncoherent03}
K. Takenaka, J. Nohara, R. Shiozaki, and S. Sugai, Phys. Rev. B {\bf 68},
  134501  (2003).

\bibitem{OrtolaniFrequency-Dependent05}
M. Ortolani, P. Calvani, and S. Lupi, Physical Review Letters {\bf 94},  067002
   (2005).

\bibitem{JakliifmmodeCharge95}
J. Jaklic, P. Prelovsek, Phys. Rev. B {\bf 52},  6903  (1995).

\bibitem{YoshidaMetallic03}
T. Yoshida {\it et~al.}, Phys. Rev. Lett. {\bf 91},  027001  (2003).

\end{thebibliography}

\end{document}